\documentclass[twocolumn,longbibliography,aps,prc,superscriptaddress,showpacs,floatfix]{revtex4-1}
\usepackage{graphics,epsfig,graphicx}
\usepackage{amsbsy,gensymb}
\usepackage{amsmath}
\usepackage{bm}
\usepackage{amsfonts}
\usepackage{cancel}
\usepackage{multirow}
\begin{document}

\title{Gaussian characterization of the unitary window for $N=3$: bound, scattering and virtual states}

\author{A. Deltuva}
\affiliation
{Institute of Theoretical Physics and Astronomy, 
Vilnius University, Saul\.etekio al. 3, LT-10257 Vilnius, Lithuania}
\author{M. Gattobigio}
\affiliation{
 Universit\'e C\^ote d'Azur, CNRS, Institut  de  Physique  de  Nice,
1361 route des Lucioles, 06560 Valbonne, France }
\author{A. Kievsky} 
\affiliation{Istituto Nazionale di Fisica Nucleare, Largo Pontecorvo 3, 56100 Pisa, Italy}
\author{M. Viviani} 
\affiliation{Istituto Nazionale di Fisica Nucleare, Largo Pontecorvo 3, 56100 Pisa, Italy}

\begin{abstract}
The three-body system inside the unitary window is studied for three equal bosons and three
equal fermions having $1/2$ spin-isospin symmetry. We perform a gaussian characterization
of the window using a gaussian potential to define trajectories for low-energy quantities
as binding energies and phase shifts. On top of this trajectories experimental
values are placed or, when not available, quantities calculated using realistic potentials 
that are known to reproduce experimental values.
The intention is to show that the gaussian characterization of the window, thought as
a contact interaction plus range corrections, captures the main low-energy properties of real
systems as for example three helium atoms or three nucleons. The mapping of real systems
on the gaussian trajectories is taken as indication
of universal behavior. The trajectories continuously link the physical points 
to the unitary limit allowing for the explanation of strong correlations between observables 
appearing in real systems and which are
known to exist in that limit. In the present study we focus on low-energy bound, scattering and virtual states.
\end{abstract}

\maketitle

%
%
%
%

\section{Introduction}
The observation of universal behavior in weakly bound systems is at present an intense
subject of research and is intimately linked to scale symmetry. One
control parameter determines all the observables in some particular energy region. 
As example we can observe that, at very low energies the scattering of two particles proceeds 
via $s$-wave and it is essentially determined
by two parameters: the two-body scattering length $a$ and the effective range $r_e$  
\begin{equation}
k \cot\delta =-\frac{1}{a}+\frac{r_e}{2} k^2 + \ldots \;\; ,
\label{eq:skk}
\end{equation}
where $E=k^2 \hbar^2/m$ is the energy of the system and $m$ is the particle mass. 
This simple relation, known as the effective range expansion~\cite{bethe}, introduces
the concept of universal behavior emanating from the scale invariance; $a$ is the control 
parameter and $r_e$ is a finite-range parameter. The details of the interaction are unimportant, 
they appear through the shape parameter in the next term of the
expansion and are unfolded as the energy increases. 
Of particular interest is the case when a shallow two-body bound state
exists close to threshold. The above expansion can be extended to locate the complex energy pole
\begin{equation}
\frac{1}{a_B}=\frac{1}{a}+\frac{r_e}{2} \frac{1}{a_B^2} \;\; ,
\label{eq:skka}
\end{equation}
where we have introduced the two-body energy length $a_B$, defined from the two-body binding
energy $E_2=\hbar^2/m a_B^2$, i.e., it is the inverse of the binding momentum $\kappa_2 = 1/a_B$.
The shallow state verifies $a_B\approx a \gg `R$ and the particles remain
much of the time outside the interaction range $R$. This condition defines the unitary window
with its central point, the unitary limit, defined by $a=\infty$ or $E_2=0$.
In the case of large and negative values of $a$ the two-body state is a virtual state.
 
Inside the unitary window the two-body system shows a continuous scale invariance (CSI). 
All two-body observables are controlled by $a$. For example $E_2$, extracted from the above equation, is
\begin{equation}
E_2=\frac{\hbar^2}{m a^2}\left(1+\frac{r_e}{a}\ldots \right ).
\label{eq:ekka}
\end{equation}
The ratio $r_e/a$ is a small parameter with the limiting case $r_e/a=0$.
In the zero-range limit we have $r_e=R=0$, and the ratio $r_e/a=0$ is always verified. In this case
the CSI is strictly verified with all the observables determined by $a$, the binding energy $E_2=\hbar^2/ma^2$
and the phase-shift $k\cot\delta=-1/a$.

The zero-range limit has been used many times to analyze the
particular structure of few-body systems inside the unitary window. In the case of an effective field theory
(EFT), the control parameter $a$ determines
the leading order (LO) of the theory whereas the range corrections appear perturbatively in the successive
terms~\cite{vankolck1,vankolck2}. Here we proceed differently based on the following observation: 
inside the unitary window the binding energy, $E=\kappa^2 \hbar^2/m$, of a generic system follows 
trajectories in the $(r_0\kappa,r_0/a_B)$ plane characterized by a length  $r_0$. 
For example the length $r_0$ can be chosen as the range of the gaussian potential
\begin{equation}
 V(r)= V_0 e^{-r^2/r_0^2} 
\label{eq:tbgaus}
\end{equation}
and such trajectories are determined by solving the corresponding Schr\"odinger equation. 
The gaussian form selected for the finite-range interaction is not special, it can be seen as a regularized
contact interaction. In this respect other representations of the delta function in the limit $r_0\rightarrow 0$
can be used as well~\cite{gatto2014}. In other words, the link from the zero-range to a finite-range
interaction is governed by one parameter for all interacting systems inside the unitary window; this
parameter is the length entering in the particular finite-range form used to characterized the unitary window.

Some aspects of the gaussian characterization of the unitary window for few-boson and fermion systems 
have been discussed in Refs.~\cite{raquel,kievsky2016, gatto2019b}.
In those works the discrete energy spectrum has been analyzed;
here we extend the discussion including in the analysis continuum states in the low-energy region paying 
particular attention to the description of two systems naturally located inside the unitary window, 
the system composed 
of three ${}^4\mathrm{He}$ atoms and of three nucleons. In both cases the ratio $r_e/a$ is smaller than one; 
in the nuclear case this is verified in both spin channels. The objective of this analysis is to
see if the gaussian characterization of the unitary window can be used to reproduce quantitatively
experimental data in order to make more clear the universal aspects of the system sometimes hidden
by finite-range effects. Furthermore, the fact that one range parameter is enough to 
classify different systems and observables extends the concept of universality. For example, many times
in the literature it has been discussed whether the ground state of the helium trimer can be considered as an
Efimov state whereas there is consensus to indicate the excited state as Efimov state~\cite{efimov1,efimov2}. 
It is clear that the latter is less affected by range corrections; 
however, we will see that through the gaussian characterization
both states behave similarly. Other observables discussed in the present analysis are the atom-dimer
scattering length and, regarding nuclear physics, we discuss the universal characterization of the
triton, its virtual state and neutron-deuteron low-energy scattering.

The paper is organized as follows. In section II we discuss the two-body system
whereas the three-body systems is discussed in section III, first for bosons and then
for fermions. In section IV the virtual states are discussed. Most of the results for
the three-body systems are obtained by two methods finding a very good
agreement between them. One is based on the Alt-Grassberger-Sandhas version of
the Faddeev equations and uses momentum-space framework; see
Ref.~\cite{deltuva:15d} and reference therein for details. The other method is
based on the expansion of the three-body wave function in terms of the hyperspherical harmonic
basis~\cite{kievsky1997,rep08}. Bound states are obtained using the Rayleigh-Ritz variational principle
whereas the Kohn variational principle is used for scattering states.
In the last section the conclusions are given.

\section{Gaussian characterization of the unitary window for two particles}

The dimensionless Schr\"odinger equation for two particles forming a $s$-wave bound state, 
interacting through a gaussian potential as given in Eq.(\ref{eq:tbgaus}), is the following
\begin{equation}
 \left(\frac{\partial^2}{\partial z^2} - \frac{mr_0^2 V_0}{\hbar^2}e^{-z^2} 
      -\frac{r_0^2}{a_B^2}\right) \phi(z) =0
\label{eq:tbsch}
\end{equation}
where $z=r/r_0$ and $\phi(z)$ is the reduced wave function. The small value of the ratio 
$r_0/a_B$ can be used to characterize the unitary window and, limiting the discussion to the
case of one bound state, the discrete energy values of all possible 
gaussians inside the window can be organized in the unique curve shown in Fig.1. In the figure
the binding momentum $\kappa_2= 1/a_B$ is given as a function of the inverse of the
scattering length $a$, both multiplied by the effective range $r_e$ to build dimensionless quantities. 
It is possible to place real systems on the gaussian curve by the corresponding experimental values of $a$, $a_B$ 
and $r_e$. To this aim we analyze two systems naturally living inside the unitary window:
the dimer of helium atoms and the two-nucleon system. For the former, instead of using
directly experimental data, which are not completely available, we use the values given by one of the widely used 
helium-helium interactions, the LM2M2 potential~\cite{lm2m2}. In the case of the two-nucleon system 
we use the experimental values of $a$ and $r_e$ in the two spin channels, the deuteron binding 
energy and the effective range expansion to determine the $nn$ and $np$ virtual states located in the 
negative $a$ region. Using the values given in Tabel I, the four cases shown in the figure by solid 
circles are on top of the gaussian curve.

\begin{table}[ht]
  \caption{Low energy quantities of the helium dimer, interacting through the LM2M2 potential,
and experimental values of the two-nucleon system in each
spin channel $S$. For the helium system the length unit is the Bohr radius $a_0$.}
  \label{tab:tbvalues}
\begin{tabular} {@{}l |  c c   c c  @{}}
\hline\hline
 helium dimer   & $\hbar^2/m\;$(K$\,a_0^2$) & $E_2\;$(mK) & $a (a_0)$ & $r_e (a_0)$     \\
                & $43.281307$           & $1.303$   & 189.415   & $13.845$        \\
\hline
 two nucleons   & $\hbar^2/m\;$(MeV$\,$fm$^2$) & $E_2\;$(MeV)  & $a\;$(fm) & $r_e\;$(fm) \\
 $np$ $S=1$     & $41.471$                   & $2.2245$    & 5.419     & 1.753       \\ 
 $np$ $S=0$     & $41.471$                   & $0.0661$    & -23.740   & 2.77        \\ 
 $nn$ $S=0$     & $41.471$                   & $0.1017$    & -18.90    & 2.75        \\ 
\hline
\end{tabular}
\end{table}

Fig.1 shows the position of different systems inside the unitary window, the helium dimer being
the nearest system to the unitary limit (the figure is useful to place different
systems inside the window). In addition deviations from the $a=a_B$ dashed line
give the size of the finite-range corrections; they are encoded in the small
parameter $r_e/a$ with values running from  $\approx 0.07$
in the case of the helium dimer to $\approx 0.32$ in the deuteron case.
In the figure the effective range has been used to make the quantities
dimensionless, however this quantity changes point by point. In Fig.2,
we reformulate the same plot in terms of the gaussian range $r_0$. 
In this type of plot real systems are mapped on the gaussian curve
through the ratio $a/a_B=\tan\theta$. 
Their positions identify the corresponding values of $r_0$ of the
gaussian potential reproducing simultaneously both $a$ and $a_B$. The vertical
lines in the figure indicate those values; we found that for the deuteron
a gaussian potential with the range $r_0=1.553\;$fm is able to reproduce simultaneously the binding
energy, the scattering length and the effective range. Similarly with a gaussian range
$r_0=10.03\;a_0$ ($a_0$ is the Bohr radius), these quantities, 
as given by the LM2M2 potential are reproduced as well.
For the $np$ and $nn$ virtual states the gaussian ranges are $r_0=1.83\;$fm and $1.795\;$fm
respectively. What we have discussed here is a gaussian characterization of the unitary
window and the identification of a range at which the gaussian potential will describe
simultaneously $a$, $a_B$ and $r_e$ and, therefore, through the effective range expansion
will describe low-energy phase shifts. We can consider these gaussian potentials a low-energy
representation of the interactions between the different systems of particles.
 
\begin{figure}[h]
\includegraphics[scale=0.34]{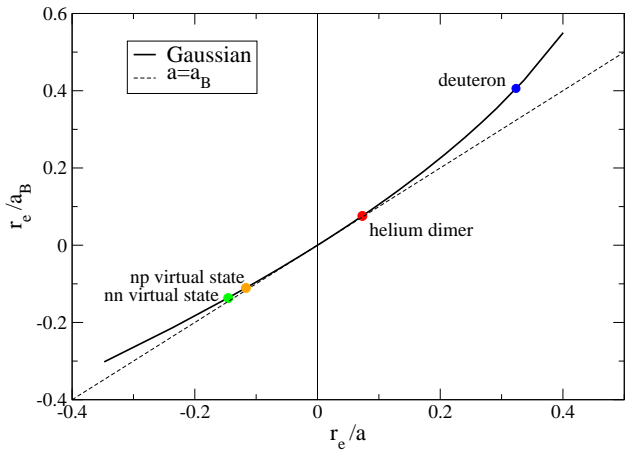}
\caption{Binding momentum as a function of the inverse scattering length for a gaussian
potential, both multiplied by the effective range $r_e$. Experimental data of four real 
systems are shown by the filled circles.}
\label{fig:gs1}
\end{figure}

\begin{figure}[h]
\includegraphics[scale=0.38]{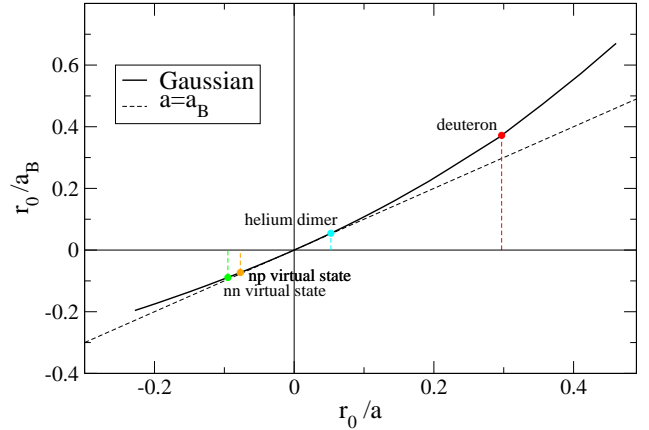}
\vspace{-0.7cm}
\caption{Binding momentum as a function of the inverse scattering length for a gaussian
potential, both multiplied by the gaussian range $r_0$. Data of real systems are located in the plot
through the ratio $a/a_B$.}
\label{fig:gs0}
\end{figure}

\section{The three-body case}

\subsection{Three equal bosons}

Inside the unitary window the three-boson $s$-wave spectrum has a
particular form as was deduced for the first time by V. Efimov~\cite{efimov1,efimov2}.
In the case of zero-range interactions the system is unbounded from
below; this property is known as the Thomas collapse~\cite{thomas}. Moreover the CSI is broken
and the residual symmetry is the discrete scale invariance (DSI).
The physics is invariant under the rescaling $r\rightarrow
\Lambda^n r$, where the constant is usually written
$\Lambda=\text{e}^{\pi/s_0}\approx 22.7$, with $s_0\approx 1.00624$ an universal number
characterizing the three-identical boson system. At the unitary limit
the spectrum consists of a geometrical series of states which accumulate at zero energy 
with ratio between two consecutive energy states $E_3^{(n+1)}/E_3^{(n)} = \text{e}^{-2\pi/s_0}$.
This effect is known as the Efimov effect. 
Many experimental efforts have been done to study this scenario using atomic traps
~\cite{kraemer2006,zaccanti2009,ferlaino2011,matchey2012,roy2013,cornell2017}.
As the system moves away from the unitary limit the highest excited states disappear into
the atom-dimer continuum one by one. A detailed analysis of three-boson spectrum for
systems having a large scattering length can be found in the reviews by Braaten and Hammer~\cite{report}
and by Naidon and Endo~\cite{naidon}. A discussion of these systems
within the EFT framework can be found in Refs.~\cite{vankolck1,vankolck2} whereas
in Ref.~\cite{gatto2019} a detailed parametrization of the zero-range universal function has been done.
When the Schr\"odinger equation is solved using a regular, finite-range, potential located at unitarity, i.e.
when there is a two-body bound state at threshold, the three-body system is bound from below 
(the Thomas collapse is not present anymore). However an infinite number of excited states
appear above the ground state showing the Efimov effect. Though there are some range effects  
in the energy ratios for the lowest states, the constant ratio of $\approx 22.7^2$ 
can be seen for ratios calculated between consecutive higher states.

We proceed to the characterization of the unitary window for three particles in the same way
as for the two-body system. We solve the Schr\"odinger equation for three equal mass bosons interacting 
through a gaussian potential. The first two levels are shown in Fig.3 where
the three-body binding momenta $\kappa^{(n)}_3=\sqrt{m|E^{(n)}_3|/\hbar^2}$ for
$n=0,1$ have been calculated as a function of the two-body
binding momentum $\kappa_2=1/a_B$, both multiplied by the gaussian range $r_0$ to build dimensionless quantities.
At the unitary limit the pure numbers
$\kappa_*^{(n)} r_0$, indicated in the figure for $n=0,1$, are the same for all gaussian
potentials. The energy goes quadratically to $\infty$ as $r_0\rightarrow 0$, a simple way to 
illustrate the Thomas collapse. The ratio
of the first two levels, $\kappa_*^{(0)}/\kappa_*^{(1)}=22.98$ shows a small range effect. If we
consider the third level, $n=2$, (not shown in the figure), we obtain 
$\kappa_*^{(2)} r_0=0.0009362$, and the ratio
$\kappa_*^{(1)}/\kappa_*^{(2)}=22.70$ is almost equal to the universal ratio.

As the system moves from the unitary limit towards the positive region the infinite
tower of excited states disappear one by one
into the $1+2$ continuum crossing the two-body threshold, shown in the plot by the solid red
curve. The second and third excited states are the last ones to cross the threshold around the
values $r_0/a_B = 0.013$ and $0.00069$, respectively. These pure numbers are the same for all gaussian potentials. 
The ground state and the first excited state do not cross the threshold; they form a two-level structure
already observed in the helium trimer~\cite{kunitski}. The two levels of the helium trimer
are located on Fig.3 (solid squares) through the angle defined by $\kappa^{(n)}_3 a_B=\tan\xi_n$. 
Using the LM2M2 potential to describe those states,
the position on the $r_0/a_B$ axis is not the same for both levels. It is 0.0612 for the
ground state and 0.0637 for the first excited state, using the values given in Table I, 
the following gaussian radius can be extracted: $r_0^{(0)}=11.15\;a_0$ and $r_0^{(1)}=11.70\;a_0$, 
respectively.  With the former range the gaussian potential describes simultaneously the dimer
and trimer ground state energies whereas with the latter the dimer and the first excited state energies.
With the values of $r_0^{(0)}$ we can predict the binding energy values of the two levels at
unitarity
\begin{eqnarray}
E_*^{(0)}=\frac{\hbar^2}{m}\left[\kappa_*^{(0)}\right]^2 = 
          \frac{\hbar^2}{m}\left[\frac{0.4883}{r_0^{(0)}}\right]^2 =83.0\; {\rm mK} \\
E_*^{(1)}=\frac{\hbar^2}{m}\left[\kappa_*^{(1)}\right]^2 = 
          \frac{\hbar^2}{m}\left[\frac{0.02125}{r_0^{(0)}}\right]^2 =0.157\; {\rm mK} 
\label{eq:unie3}
\end{eqnarray}
to be in a good agreement with the values given in the literature for the LM2M2 
potential when its strength is reduced in order to locate the two-body bound
state at threshold~\cite{barletta,hiyama2014}. 

\begin{figure}[h]
\includegraphics[scale=0.34]{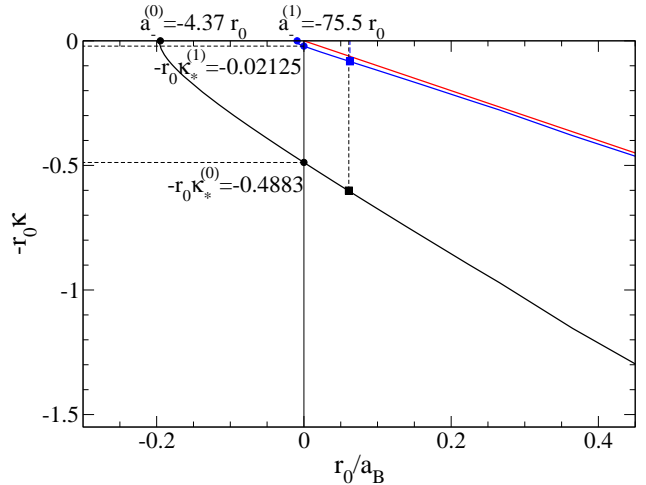}
\caption{Three-body binding momentum as a function of the two-body binding momentum
for a gaussian potential, both multiplied by the gaussian range $r_0$ to build
dimensionless quantities. Notable points are indicated as solid circles
as explained in the text.}
\label{fig:bosong}
\end{figure}

Decreasing the strength of the gaussian potential the system enters the region of negative $a$
in which two particles are not bound. In this case $a_B$ is related to the energy of the two-body
virtual state. At some point the first excited state of the trimer disappears into
the three-atom continuum. Decreasing further the strength the interaction cannot bind the three
particles and the ground state disappears too. The two-body
scattering lengths at which these transitions occurs, indicated as $a_-^{(1)}$ and $a_-^{(0)}$ 
respectively, are shown by solid circles on top of Fig.3 in units of
the gaussian range. Using the characteristic gaussian range,
$r^{(n)}_0$, we can predict the values of the scattering length at threshold for
the LM2M2 potential as $a_-^{(0)}\approx -48.7\;a_0$  and
$a_-^{(1)}\approx -842\;a_0$ to be compared to $a_-^{(0)}\approx -48.2\;a_0$ and
$a_-^{(1)}\approx -832\;a_0$ given in Ref.~\cite{hiyama2014}. 

This analysis demonstrates that the gaussian characterization of the states
around the unitary window captures the essential ingredients of the dynamics. A more stringent test
beyond the study of helium trimers is given by the almost model independent product
$a_-^{(0)}\kappa_*^{(0)}\approx -2.2 $ obtained through an analysis of experimental data
for different van der Waals potentials having a repulsive core and supporting one or more bound
states (see Ref.\cite{naidon} and references therein).
Surprisingly the gaussian potential predicts the values
\begin{eqnarray}
a_-^{(0)}\kappa_*^{(0)}= -2.14  \\
a_-^{(1)}\kappa_*^{(1)}= -1.56  
\end{eqnarray}
capturing properties related to the van der Waals tail in the case of ground state level. 
In the case of the first excited state the product $a_-^{(1)}\kappa_*^{(1)}$ is
close to the product obtained in zero-range limit ($a_-\kappa_*=1.507$, see
Ref.~\cite{aminus}) showing the minor size of the finite-range corrections in this state. 
The analysis of the $n=2$ level gives results comparable to the zero-range
corrections at the $0.1\%$ level or better~\cite{raquel}.
Within the EFT framework the results for the ground state
can be considered at the level of next-to-leading order (NLO) whereas those for the excited state 
at LO level.

Another observable we can study using the gaussian characterization of the universal window is 
the atom-dimer scattering length. As derived by Efimov~\cite{efimov3}, this observable has a 
universal expression in the zero-range limit
\begin{equation}
a_{AD}/a_B=d_1 +d_2\tan[s_0\ln(\kappa_*a_B)+d_3]\,,
\label{eq:a_ADB}
\end{equation}
where $d_1$, $d_2$ and $d_3$ are universal numbers and $\kappa_*$ is the three-body
parameter labeling one three-body branch. This is a particular realization of the DSI with
the log-periodic functional form and accordingly, as $a_B\rightarrow\infty$, the ratio $a_{AD}/a_B$ repeats its
values forming different branches with asymptotes in the points in which the excited states disappear
into the atom-dimer continuum. It should be noticed that a numerical analysis of the above form is very delicate
due to the extremely weak binding of the dimer as $a_B\rightarrow\infty$. Here
we analyze the behavior of $a_{AD}$ inside the unitary window for a gaussian potential up to the
region in which the third excited state is present. The
results are given in Fig.4 where much of the results presented in this section are collected.
The (blue) solid circles are the numerical results of $a_{AD}/a_B$, obtained solving
the zero-energy scattering problem, as a function of $r_0/a_B$. Presented in this way the results are 
the same for all gaussian potentials at the same $r_0/a_B$ value.
The (blue) solid line has been obtained using the following parametrization 
of the function
\begin{equation}
a_{AD}/a_B=d_1 +d_2\tan[s_0\ln(\kappa^{(1)}_*r_0(a_B/r_0)+\Gamma_3^{(1)})+d_3]\,,
\label{eq:a_ADR}
\end{equation}
where we have used the pure number, $\kappa^{(1)}_* r_0=0.02125$, as the driving
term and we have introduced the corresponding shift $\Gamma_3^{(1)}$, as discussed in Ref.~\cite{gatto2013}.
A reasonable fit to the calculations is obtained with $d_1=1.541$, $d_2=-2.080$, $d_3=-2.038$ and 
$\Gamma_3^{(1)}=0.061$. The two vertical dashed lines, located at $r_0/a_B=0.013$ and $0.00069$,
are the asymptotes indicating the values
at which the second and third excited states disappear into the atom-dimer threshold, respectively. From the
figure we can see that the gaussian characterization of the $a_{AD}/a_B$ function has the log-periodic form
in which the finite-range corrections have been absorbed in the shift parameter $\Gamma_3^{(1)}$.

In Fig.4 the position of the excited state of the trimer (lower red diamond) on
the $n=1$ level, as given by the LM2M2 potential, is shown. Extending the corresponding value 
of the axis $r_0/a_B=0.0637$ to cross the $a_{AD}/a_B$ curve the value $a_{AD}/a_B=1.19$ is extracted.
Therefore the
gaussian characterization of the unitary window indicates, for that value of the $r_0/a_B$ ratio,
the atom-dimer scattering length to be $a_{AD}=1.19\, a_B$. 
Using the LM2M2 value, $a_B=182.22\, a_0$, we obtain $a_{AD}=217 \, a_0$, to be compared
to the LM2M2 value for this quantity of $218.4\, a_0$~\cite{carbonell}. So, a gaussian
potential constructed to describe simultaneously the dimer and excited state of the trimer
reproduces also the value of the atom-dimer scattering length within a $1\%$ error. 

To analyse further the size of the finite-range effects the numerical value of $a_{AD}$ 
can be extracted from the higher branch of the $a_{AD}/a_B$ curve, between the two 
asymptotes shown in the Fig.4. 
To this aim the curve has to be evaluated at the coordinate $r_0/a_B$ corresponding
to the position of trimer excited state on the third energy level ($n=2$). That
point has coordinates $(r_0\kappa,r_0/a_B)$ verifying $\kappa\, a_B=\tan\xi_1$,
the same angle of the point on the $n=1$ level, indicated by the lower red
diamond on the figure. The coordinate is $r_0/a_B=0.000239$ and corresponds to the value $a_{AD}/a_B=1.17$ 
giving $a_{AD}\approx 213 \, a_0$. 
As for bound states, the result obtained using the lower branch can be considered at
NLO level whereas using the higher branch at LO level. 

This analysis shows that, inside the unitary window and in the low energy region, the complicate 
structure of the helium potential can be encoded in the strength and range of the a gaussian potential. 
The lowest energy state or branch, as in the case of atom-dimer scattering, captures the essential 
elements of system and can be
considered at the NLO level in the EFT framework. The size of the corrections can be
evaluated from the analysis of the higher states or branches and give results at the LO level.
Moreover, varying the gaussian parameters a complete picture of the unitary window can be depicted.

\begin{figure}[h]
\includegraphics[scale=0.34]{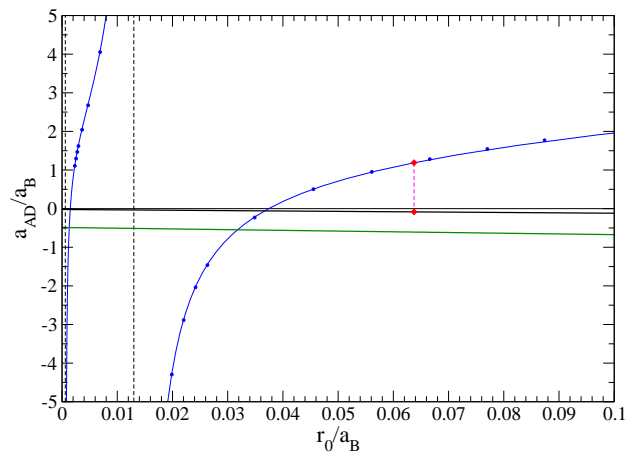}
\caption{The atom-dimer scattering length (in units of $a_B$) as a function of the dimer binding momentum
 $\kappa_2=1/a_B$ times $r_0$. The blue solid points are the calculations using a 
gaussian potential and the blue solid line is the parametrization of Eq.(\ref{eq:a_ADR}).
The binding momentum (times $r_0$) of the ground state (green solid line) and the first
excited state (black solid line) are shown too. The solid red diamonds indicate the positions of the
first excited state of the helium trimer and the atom-dimer scattering length. The two dashed lines indicate
the position in which the second and third excited states cross the atom-dimer threshold. }
\label{fig:fig4}
\end{figure}

\subsection{Three $1/2$ spin-isospin fermions}

We perform the gaussian characterization of the unitary window focusing on the
three-nucleon system. The intention is to link the triton continuously to the
unitary limit showing that particular characteristics of the three-nucleon system in
the low-energy region are strictly related to its position in the unitary window.
In particular we would like to explain the one-level structure of the triton, the
almost zero value of the doublet neutron-deuteron scattering length $a_{nd}$, and the position
of the triton virtual state indirectly observed through the $s$-wave phase shifts
in low-energy neutron-deuteron scattering. Studies of the three-nucleon systems inside
the unitary window can be found in Refs.~\cite{kievsky2016,gatto2019b,koenig,kievsky2017,kievsky2018}.

To perform the gaussian characterization of the unitary window for the three-nucleon
system we use a spin dependent gaussian interaction with different terms in both
spin channels $S=0,1$  
\begin{equation}
 V(r)= V_0 e^{-r^2/r_0^2} {\cal P}_0+V_1 e^{-r^2/r_1^2} {\cal P}_1
\label{eq:tbgaussd}
\end{equation}
where ${\cal P}_0$ and ${\cal P}_1$ are the projectors onto the spin-isospin channels
$S,T=0,1$ and $1,0$ respectively. Moreover, we limit the two-body force to act in $s$-waves.
The nuclear force is weak in 
angular momentum states $\ell >0$ and, in the low-energy region  considered here,
this restriction could be justified. In the study of the unitary window with such a force
we consider the two gaussian ranges to be equal $r_0=r_1$. In this case, when
$V_0$ is equal to $V_1$, the spectrum of total angular momentum and
parity $J^\pi=1/2^+$ states is equivalent to the three-boson spectrum discussed previously.

Among different possibilities to characterize the unitary window, following
Refs.~\cite{kievsky2016,gatto2019b}, we select trajectories
maintaining constant the ratio of the singlet and triplet scattering lengths, $a_0$ and $a_1$, 
and equal to the nuclear physics value $a_0/a_1\approx -4.3$. With this condition,
we vary the gaussian parameters
to cover the plane $(-r_0 \kappa_3, r_0/a_B)$, with $E_3=\hbar^2 \kappa^2_3/m$ the three-body binding energy
of the $J^\pi = 1/2^+$ state and $E_2= \hbar^2/ma_B^2$ the two-body binding energy of the triplet
state. The results are shown in the two panels of Fig.5 for discrete states and the zero-energy solution. 
The upper panel shows the 
$nd$ doublet scattering length ${}^2a_{nd}$, in units of the energy length
$a_B$, as a function of $r_0/a_B$. In the negative 
region we show the binding momenta (times $r_0$) for the ground state (orange curve) and excited state (green curve) of the trimer. 
The dashed vertical line is the asymptote, at $r_0/a_B=0.101$, indicating the point in
which ${}^2a_{nd}$ diverges and the trimer excited state disappears into the $1+2$ continuum.
After the asymptote, the red curve shows the binding momentum of dimer (times $r_0$), in the $S=1$ state.
The physical point, indicated by the red diamond on the trimer curve, has
$\kappa_3 a_B=\tan\xi=1.95$ 
corresponding to the
square root of the ratio of the triton binding energy $8.48\,$MeV with the deuteron
binding energy of $2.224\,$MeV. At that point $r_0/a_B=0.457$ and ${}^2a_{nd}/a_B=0.08$ allowing to extract
the value of ${}^2a_{nd}=0.4\,$fm. This value is slightly lower of the experimental
value ${}^2a_{nd}\approx0.65\,$fm. However, the gaussian characterization is able to explain
the almost zero value of this quantity if compared to the triplet $np$ scattering length $a_1=5.42\,$fm. 
This is a very delicate region in which slightly different values of $r_0/a_B$  produce large variations of
${}^2a_{nd}$, including a change of sign.
The gaussian characterization places ${}^2a_{nd}$ in the correct (positive) 
region and shows the strong correlation
between this quantity and the trimer energy observed already many years ago and known as the
Phillips line~\cite{phillips}. As for the boson case it is possible to analyse
the higher branch of the ${}^2a_{nd}/a_B$ curve to determine the size of
finite-range corrections. To this aim the triton point is located on the $n=1$
level (green curve on the upper panel of Fig.5) at coordinates having the same
value $\tan\xi=1.95$, this happens at $r_0/a_B=0.015$. At that coordinate
${}^2a_{nd}/a_B=0.06$, slightly lower than the value obtained analysing
the $n=0$ level.

The lower panel of Fig. 5 shows calculations of the doublet scattering length ${}^2a_{nd}$ (solid
blue circles) in a extended region and a fit to those values using the form of Eq.(\ref{eq:a_ADR}) 
(solid blue curve) with $d_1=0.542$, $d_2=-1.686$, $d_3=6.1952$, the shift $\Gamma_3^{(0)}=0.378$ and
the driving term of the ground state energy $\kappa_* r_0=0.488$. Using the fit we can
extract the asymptote at $r_0/a_B=0.101$ and determine the energy length at which the excited
state disappears $a_B\approx 16\,$fm. This analysis explains the existence of one bound
state for the three-nucleon system: at the physical values of $a_0$ and $a_1$ the excited
state has crossed the threshold becoming a virtual state. Finally, using the value
$r_0/a_B=0.457$ at the triton point and from the deuteron length, $a_B=4.32\,$fm,
we can extract the characteristic gaussian range $r_0=1.97\,$fm from which it is possible
to assign a value of the three-nucleon system at unitary through the quantity
$\kappa_* r_0=0.488$. We obtain $E_*\approx 2.5\,$MeV in good agreement with previous 
estimates~\cite{gatto2019b,epelbaum}.

\begin{figure}[h]
\includegraphics[scale=0.34]{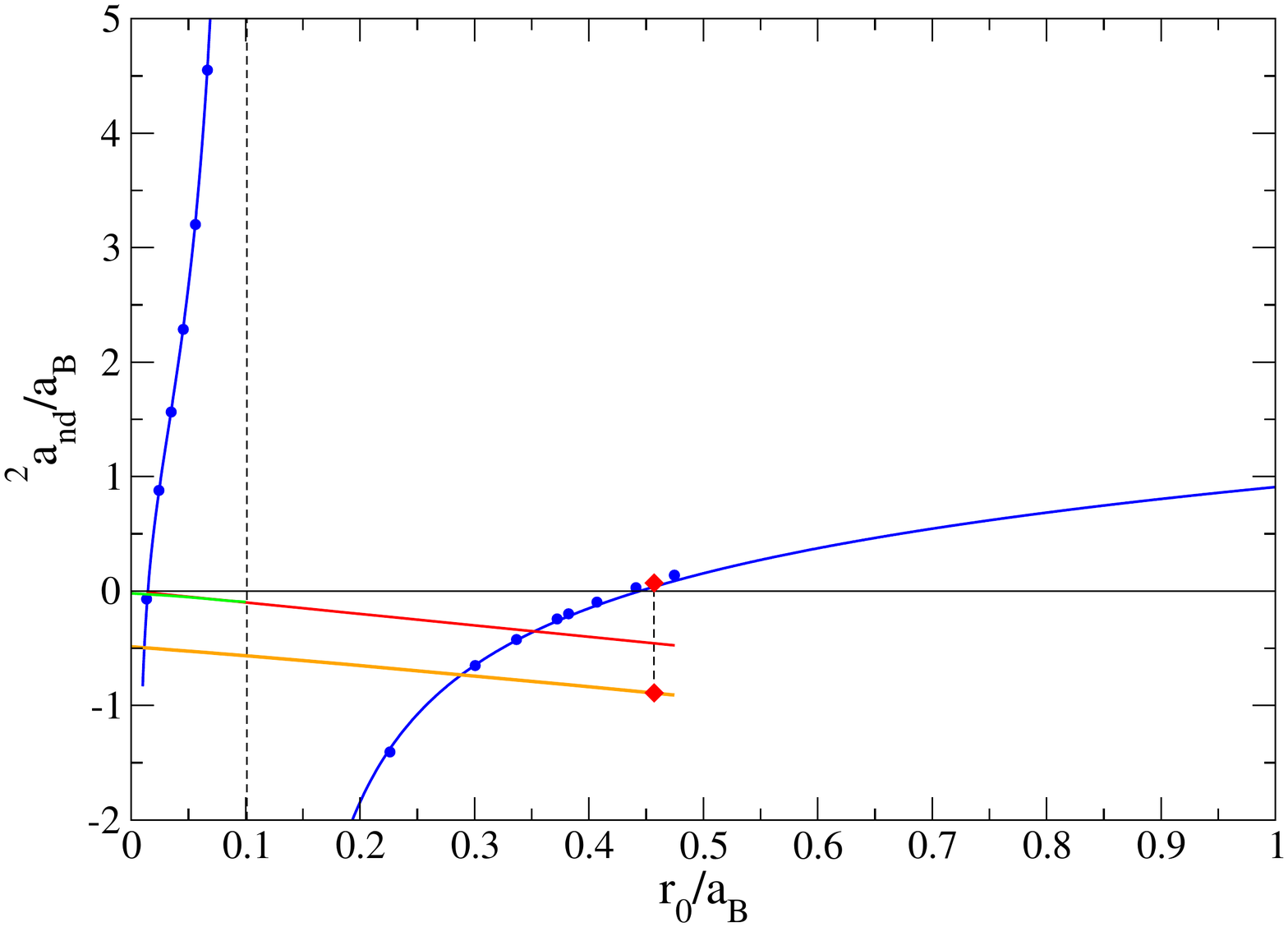}
\includegraphics[scale=0.34]{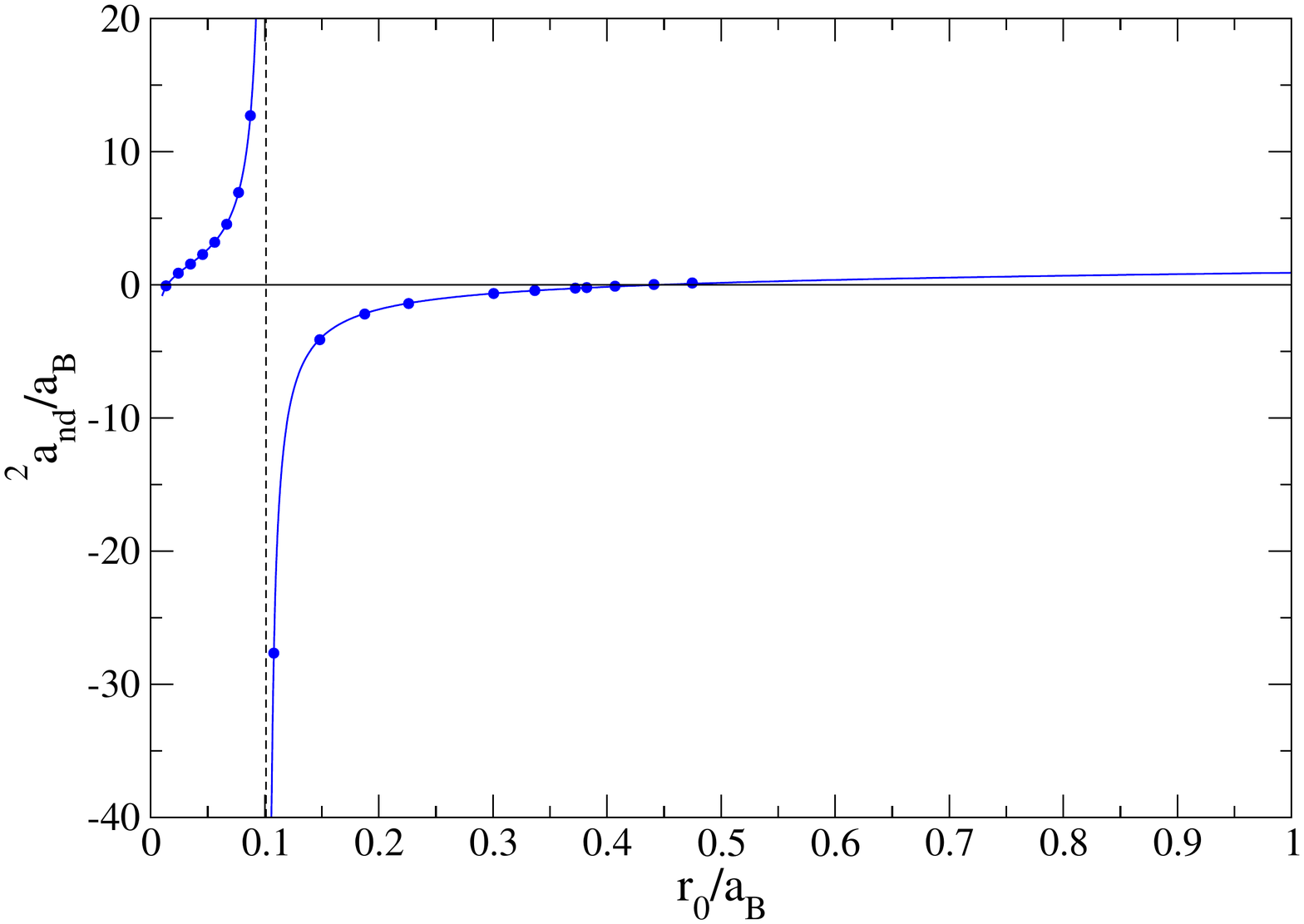}
\caption{The doublet neutron-deuteron scattering length ${}^2a_{nd}$ in units of the
triplet two-body energy length $a_B$ as a function of the inverse of $a_B$ in units
of the gaussian range $r_0$ (blue solid points). The blue line is the parametrization
of Eq.(\ref{eq:a_ADR}).
In the upper panel the dimensionless binding momenta $\kappa^{(n)}_3 r_0$ of the
three-nucleon ground state (orange curve), $n=0$, and excited state (green curve), $n=1$,
are shown together with the binding momentum $\kappa_2 r_0$ of the two-body bound state (red curve).
 The excited state almost overlaps with the two-body state. The red diamonds 
indicate the physical point of the ground state curve and the corresponding value on the
${}^2a_{nd}$ curve. The lower panel shows ${}^2a_{nd}/a_B$ in a larger region. The vertical dashed line is the
asymptote at which ${}^2a_{nd}$ diverges.}
\label{fig:fig5}
\end{figure}

\section{The three-body virtual states}

We have studied the three-body energy spectrum inside the unitary window using a gaussian
interaction with variable strength and we have observed that
when the excited state cross the $1+2$ threshold the particle-dimer
scattering length diverges and the excited state becomes a virtual state. 
Here we study the evolution of the virtual
state after that crossing. For bosons we saw that the first excited state never crosses the
threshold, the last level to cross the threshold is the second excited state 
quite close to the unitary limit and therefore it  has little effects in 
 low-energy atom-dimer collisions as the system moves away from the unitary
limit. Conversely, we have observed that for fermions along the nuclear trajectory the first
excited state crosses the threshold becoming a virtual state.
As we will see, its position at the physical point could be determined
from the low-energy neutron-deuteron scattering. To this respect, the virtual state of the triton has been
subject of different studies for a long time, see Refs.~\cite{fuda,babenko}, for a recent
review see Ref.~\cite{orlov} and references therein whereas in Ref.~\cite{higa} a treatment within EFT is 
presented.

In order to determine the position of the virtual state we calculate
the $s$-wave effective range function $S_k=k\cot\delta$, where $k$ is defined from the
center of mass particle-dimer energy $E=3\hbar^2 k^2/4m$ and $\delta$ is the corresponding
phase shift. In the case of bosons
\begin{equation}
\lim_{k\rightarrow 0} S^B_k = -\frac{1}{a_{AD}}.
\end{equation}
In the case of $1/2$ spin-isospin
fermions we study neutron-deuteron scattering in the $J^\pi = 1/2^+$ state 
with
\begin{equation}
\lim_{k\rightarrow 0} S^F_k = -\frac{1}{{}^2a_{nd}}.
\end{equation}
The superscripts $B$ or $F$ in the effective range function identify the different symmetries.
The effective range functions, calculated using a gaussian potential, are given in Fig.6 for different values 
of center of mass energy energies.
In the figure the dimensionless quantities
$a_{AD} S^B_k$ (upper panel) and ${}^2a_{nd} S^F_k$ (lower panel), for bosons and fermions, respectively,
are shown as a function of $a_B k$. Using this form
both functions start at $-1$ at zero energy and the breakup threshold is $a_Bk=2/\sqrt{3}$. 
The different curves are labelled by the ratio $r_0/a_B$
and correspond to some of the $a_{AD}$ and ${}^2a_{nd}$ values given in Fig.4
and Fig.5. From the effective
range function, or directly from the phase-shifts at different energies, it is possible to construct a
representation of the $S-$matrix and extract the pole located at the negative imaginary axis
(the virtual state). 

\begin{figure}[h]
\begin{center}
 \includegraphics[scale=0.45]{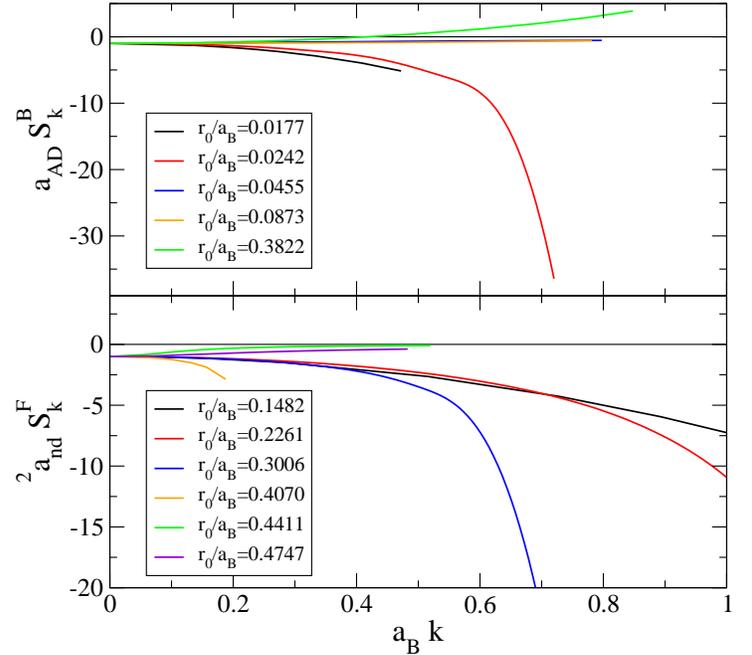}
\end{center}
\caption{The effective range functions $S^B_k$ (upper panel) and $S^F_k$ (lower panel)
as functions of the energy momentum $k$, multiplied by energy length $a_B$.
The different curves correspond to different dimer binding energies.}
\label{fig:fig6}
\end{figure}

Close to the $1+2$ threshold the virtual state can be detected through a particular
form of the effective range function. Following Ref.~\cite{orlov}, the effective
range function can be parameterized as
\begin{equation}
S_k^\lambda \approx \frac{-1/a_\lambda+C^\lambda_2k^2+C^\lambda_4k^4}{1+k^2/(k^\lambda_0)^2}
\end{equation}
with $\lambda=B,F$ and $a_\lambda,C^\lambda_2,C^\lambda_4,k^\lambda_0$ parameters used to 
fit experimental values for phase-shifts
or, when not available, numerical results using potential models.
As discussed in Ref.~\cite{orlov}, these four quantities  are related to the
energy of the $S$-matrix pole, $E_p=-3\hbar^2 k^2_p/4m$. One possibility to extract the virtual state
in the unitary window using numerical results obtained with the
gaussian potential would be to determine those four parameters. We found more convenient to
use the gaussian phase-shifts directly to derive the Pad\'e approximant representation of the $S$-matrix
and extract the pole on the negative imaginary axis using the method given in Ref.~\cite{elander}.
We have followed this technique and the quantities, $a_{AD} k_p$ and ${}^2a_{nd} k_p$,
for three bosons or three-fermions along the nuclear plane, are given in Fig.7 as functions of
$r_0/a_B$. In the boson case the pole has initially an almost linear behavior and moves away from the threshold.
Conversely, in the three-nucleon case, the pole move smoothly and remains close to the threshold
even at the physical point. This allows the possibility of its determination analyzing experimental
data for $nd$ scattering close to threshold.

The position of the virtual state at the physical point for three-nucleons can be compared
to values obtained from experimental results, which are not exhaustive, or to results obtained from
realistic potential models. To analyse the latter possibility, we perform calculations
for the three-nucleon system using two realistic forces, the CD Bonn potential including the
$\Delta$-isobar excitation and the three-body force (CD-Bonn+D+U3) \cite{deltuva:15d} 
and the AV18 potential with the Urbana IX force slightly modified
to reproduce the triton binding energy and ${}^2a_{nd}$ (AV18+URIX') \cite{kievsky2010}. Furthermore the
spin-dependent two-body gaussian potential supplemented with a hyperradial 
three-body force (2BG+3BG) of the following form has also been used
\begin{equation} 
 V(i,j)+W(i,j,k)= V_0 e^{-r^2/r_0^2} {\cal P}_0+V_1 e^{-r^2/r_1^2} {\cal P}_1
                  + W_0 e^{-\rho^2/\rho_0^2}
\end{equation} 
with strength parameters $V_0=37.9\,$ MeV, $V_1=60.575\,$ MeV, $W_0=2.7947\,$MeV
and range parameters $r_0=r_1=1.65\,$fm and $\rho_0=5.05\,$fm. With this selection the
two- and three-body low-energy quantities are well described including the triton
binding energy and ${}^2a_{nd}$. The nuclear physics point corresponds to $r_0/a_B=0.457$ 
and the results of these realistic models lie almost on the gaussian curve.
For the three models, the $s$-wave  $J^\pi=1/2^+$ low-energy phase shifts
 are shown in Fig.8 from which the energy of the pole $E_p=0.48\,$MeV can
be extracted. The three pole energy values are indicated in Fig.7. As can be seen from the
figure, the gaussian characterization describes correctly this state.

\begin{figure}[h]
\includegraphics[scale=0.34]{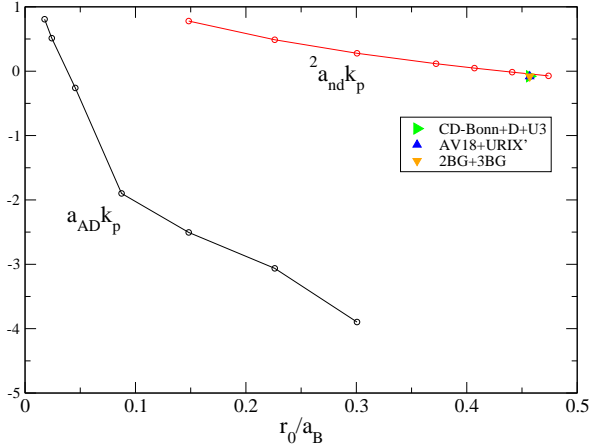}
\caption{The virtual state pole momentum for three bosons and three fermions for different
dimer energies. The realistic cases for fermions are on top of the gaussian
characteristic curve.}
\label{fig:fig7}
\end{figure}

\begin{figure}[h]
\includegraphics[scale=0.34]{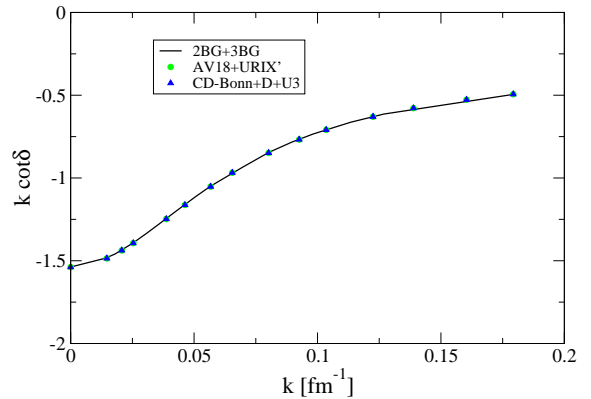}
\caption{The effective range function for the three realistic cases discussed.}
\label{fig:fig8}
\end{figure}

\section{Conclusions}

Due to the independence on the interaction details, few-body systems have been studied inside 
the unitary window interacting through a gaussian potential. This research follows
other studies that can be found in Refs.~\cite{kievsky2014,bazak,artur,artur1}. 
In the present study we perform a step further
constructing a gaussian characterization of the unitary window for different
observables in the three-body sector. The aim of this study is twofold; from one side
we would like to extend the characterization of the unitary window from the zero-range
theory to finite range. One the other side we would like to use the gaussian trajectories 
to analyze universal behavior. 
Many properties of real systems are well described using detailed interaction models, we refer
for example to realistic He-He or $NN$ interactions. Since these interactions are determined
from a set of experimental data, when a system is forced to move from its phyiscal point
the original interaction is modified in a certain (unknown) form. Broad Feshbach resonances
used to explore the unitary window are compatible with an overall strength multiplication
of the atomic potential. Essentially this kind of process is well represented by a
gaussian potential with variable strength and fixed range.
If the evolution of a system inside the unitary window follows the gaussian
trajectories we identify this system as a representative of the universal class
of weakly bound systems; in fact, a unique gaussian
range can be identified and used to accommodate observables, as binding energies,
on top of the gaussian trajectories. In this way very different systems have
been mapped on the same curve identifying their universal behavior. To this respect the
gaussian trajectories make one step further with respect to the zero-range trajectories
since they include range corrections.

Firstly, following the above methodology, we have studied the two-body system; we have 
placed on the gaussian curve four different systems, the helium dimer, the deuteron and the
$S=0$ $np$ and $nn$ systems. The position on the curve can be
used as a measure of the distance of the physical point from the unitary limit. 
Then, we have characterized the three-body system inside the unitary window; in the case of
three-equal bosons, focusing on the system of three helium atoms, and in the
case of three $1/2$ spin-isospin fermions, 
focusing on the three-nucleon system. Using a gaussian potential we have studied the energy 
spectrum, bound and excited states, and $s$-wave low-energy scattering. The next
step has been to map on the gaussian trajectories the physical points and to identify correlations
between the observables.
These correlations, used as a signal of universality, has been found to be very effective. From the
location of the physical points we have been able to predict the values of the corresponding $1+2$
scattering lengths. The helium atom-dimer scattering length has been predicted with a good accuracy.
In the case of the doublet neutron-deuteron scattering length an approximate
value has been extracted, however 
the correct position close to zero has been correctly predicted. In the former case, the almost independent
product $a^{(0)}_-\kappa_*^{(0)}$ has been found to coincide closely to the value of different
van der Waals species, giving a further confirmation of the potentialities of using gaussian
trajectories to identify universal behavior. Using the (approximate) DSI we have
analysed the higher energy levels or branches, in the case of atom-dimer or
deuteron-neutron scattering length, to assess the size of finite-range
corrections. In the framework of EFT, the results obtained from the lowest energy level or branch 
can be considered at NLO whereas the higher ones, very close to the zero-range interaction model,
represent LO results. 

In the final part we have studied the virtual states appearing when the excited states cross the
$1+2$ threshold. This is particularly interesting in the three-nucleon system where
the virtual state of the triton has been subject of many investigations. From our analysis the virtual
state is located at $E_p=-0.48\,$ MeV, consistently with previous determination~\cite{orlov}. 
The analysis presented in our work is useful to characterize low-energy properties of 
few-nucleon systems as belonging to the universal window. In conclusion, when a system is located
inside the unitary window, the ratio of the three- and two-body energies determines the
angle from which the system is placed on the trajectory and determines the characteristic gaussian
range. Then the gaussian potential with that range can be used to perform a complete characterization
of the universal window for that system. Other observables strictly correlated to the
binding energy can be predicted as well. In this way we can study scale symmetries observed in the real
systems as continuously linked to the unitary point, a point in which those symmetries are well 
verified~\cite{bira2017}.


\begin{thebibliography}{10}
\bibitem{bethe} H.A. Bethe, Phys. Rev. {\bf 76}, 38 (1949)
\bibitem{vankolck1} P.F Bedaque, H.-W. Hammer, and U. van Kolck, Phys. Rev.
Lett. {\bf 82}, 463 (1999).
\bibitem{vankolck2} P. Bedaque, H.-W. Hammer, and U. van Kolck, Nucl. Phys.
A {\bf 676}, 357 (2000).
\bibitem{gatto2014} M. Gattobigio and A. Kievsky, Phys. Rev A {\bf 90}, 012502 (2014).
\bibitem{raquel} R. \'Alvarez-Rodr\'\i guez, A. Deltuva, M. Gattobigio and A. Kievsky,
Phys. Rev. A {\bf 93}, 062701 (2016).
\bibitem{kievsky2016} A. Kievsky and M. Gattobigio, Few-Body Syst. {\bf 57}, 217 (2016)  
\bibitem{gatto2019b} M. Gattobigio, A. Kievsky, and M. Viviani,
Phys. Rev. C {\bf 100}, 034004 (2019).
\bibitem{efimov1}V. Efimov, Phys. Leet. B {\bf 33}, 563 (1970)
\bibitem{efimov2}V. Efimov, Yad. Fiz. {\bf 12}, 1080 (1970) [Sov. J. Nucl. Phys. {\bf 12}, 589 (1971)]
\bibitem{deltuva:15d} A. Deltuva and P. U. Sauer, Phys. Rev. C 91, 034002 (2015) 
\bibitem{kievsky1997}A. Kievsky, Nucl. Phys. A{\bf 624}, 125 (1997)
\bibitem{rep08} A.~Kievsky, S.~Rosati, M.~Viviani, L.~E.~Marcucci and L.~Girlanda,
, J. Phys. G: Nucl. Part. Phys. {\bf 35}, 063101  (2008)
\bibitem{lm2m2}R.A. Aziz and M.J. Slaman, J. Chem. Phys. {\bf 94}, 8047 (1991)
\bibitem{thomas} L.H. Thomas, Phys. Rev. {\bf 47}, 903 (1935)
\bibitem{kraemer2006} T. Kraemer {\sl et al.}, Nature {\bf 440}, 315 (2006).
\bibitem{zaccanti2009} M. Zaccanti {\sl et al.}, Nat. Phys. {\bf 5}, 586 (2009).
\bibitem{ferlaino2011} F. Ferlaino {\sl et al.}, Few-Body Syst. {\bf 51}, 113 (2011).
\bibitem{matchey2012} O. Matchey, Z. Shotan, N. Gross, and L. Khaykovich,
Phys. Rev Lett. {\bf 108}, 210406 (2012).
\bibitem{roy2013} S. Roy {\sl et al.}, Phys. Rev Lett. {\bf 111}, 053202 (2013).
\bibitem{cornell2017} C. E. Klauss {\sl et al.}, Phys. Rev Lett. {\bf 119}, 143401 (2017).
\bibitem{report} E. Braaten and H.-W. Hammer, Phys. Rep. {\bf 428}, 259 (2006).
\bibitem{naidon}P. Naidon and S. Endo Rep. Prog. Phys. {\bf 80} 056001 (2017)
\bibitem{gatto2019}M. Gattobigio, M. G\"obel, H.-W. Hammer and A. Kievsky,
Few-Body Syst. {\bf 60}, 40 (2019)
\bibitem{kunitski} M. Kunitski et al., Science {\bf 348}, 551 (2015).
\bibitem{barletta} P. Barletta and A. Kievsky, Phys. Rev. A {\bf 64}, 042514 (2001).
\bibitem{hiyama2014} E. Hiyama and M. Kamimura, Phys. Rev A {\bf 90}, 052514 (2014)
\bibitem{aminus} A.~O.~Gogolin, C.~Mora and R.~Egger,
  Phys.\ Rev.\ Lett.\ {\bf 100}, 140404 (2008).
\bibitem{efimov3}V. Efimov, Yad. Fiz. {\bf 29}, 1058 (1979) [Sov. J. Nucl. Phys. {\bf 29}, 546 (1979)]
\bibitem{gatto2013} A. Kievsky and M. Gattobigio, Phys. Rev. A {\bf 87}, 052719 (2013).
\bibitem{carbonell} J. Carbonell, A. Deltuva, and R. Lazauskas, Comp. Rend. Phys. {\bf 12}, 47 (2011)
\bibitem{phillips} A.C. Phillips, Nucl. Phys. A \textbf{107}, 209 (1968)
\bibitem{epelbaum} E. Epelbaum, H.-W. Hammer, U.-G. Mei\ss ner, and A. Nogga, Eur. Phys. J. C{\bf 48},
        169 (2006)
\bibitem{fuda} B.A. Girard and M.G. Fuda, Phys. Rev. C {\bf 19}, 579 (1979)
\bibitem{babenko} V.A. Babenko and N.M. Petrov Yad. Fiz. {\bf 63}, 1798 (2000) [Phys. At. Nucl. {\bf 63},
 1709 (2000)]
\bibitem{orlov} Yu.V. Orlov and L.I. Nikita, Yad. Fiz. {\bf 69}, 631 (2006) 
         [Phys. At. Nucl. {\bf 69}, 607 (2005)] 
\bibitem{higa} G.Rupak, A. Vaghani, R. Higa, and U. van Kolck, Phys. Lett. B {\bf 791}, 414 (2019)
\bibitem{elander}S.A. Rakityansky, S.A. Sofianos, and N. Elander, J. Phys. A: Math. Theor. {\bf 40}, 14857 (2007)
 \bibitem{kievsky2010}A. Kievsky, M. Viviani, L. Girlanda, and L.E. Marcucci, Phys. Rev. C{\bf 81}, 044003 (2010)
\bibitem{koenig} S. K\"onig, H.W. Grie\ss hammer, H.-W. Hammer, and U. van Kolck,
        Phys. Rev. Lett. {\bf 118}, 202501 (2017).
\bibitem{kievsky2017}A. Kievsky, M. Viviani, M. Gattobigio, and L. Girlanda,
        Phys. Rev. C{\bf 95}, 024001 (2017)
\bibitem{kievsky2018}A. Kievsky, M. Viviani, D. Logoteta, I. Bombaci, and L. Girlanda,
        Phys. Rev. Lett. {\bf 121}, 072701 (2018)
\bibitem{kievsky2014}A. Kievsky, N.K. Timofeyuk, and M. Gattobigio, Phys. Rev, {\bf A90}, 032504 (2014).
\bibitem{bazak} B. Bazak, M. Eliyahu, and U. van Kolck, Phys. Rev. {\bf A94}, 052502 (2016)
\bibitem{artur} A. Kievsky, A. Polls, B. Juli\'a D\'\i az, and N.K. Timofeyuk, 
         Phys. Rev A {\bf 96}, 040501(R) (2017)
\bibitem{artur1} A. Kievsky, A. Polls, B. Juli\'a D\'\i az, N.K. Timofeyuk, and M. Gattobigio, 
submitted for publication.
\bibitem{bira2017} U. van Kolck, Few-Body Syst. {\bf 58}, 112 (2017)


\end{thebibliography}
\end{document}